\RequirePackage{fix-cm}
\documentclass[twocolumn]{svjour3}           
\smartqed  
\usepackage{graphicx}
\usepackage{url}
\usepackage{caption}
\usepackage{subfigure}
\usepackage{scalerel}
\usepackage{tikz}
\usetikzlibrary{svg.path}

\definecolor{orcidlogocol}{HTML}{A6CE39}
\tikzset{
  orcidlogo/.pic={
    \fill[orcidlogocol] svg{M256,128c0,70.7-57.3,128-128,128C57.3,256,0,198.7,0,128C0,57.3,57.3,0,128,0C198.7,0,256,57.3,256,128z};
    \fill[white] svg{M86.3,186.2H70.9V79.1h15.4v48.4V186.2z}
                 svg{M108.9,79.1h41.6c39.6,0,57,28.3,57,53.6c0,27.5-21.5,53.6-56.8,53.6h-41.8V79.1z M124.3,172.4h24.5c34.9,0,42.9-26.5,42.9-39.7c0-21.5-13.7-39.7-43.7-39.7h-23.7V172.4z}
                 svg{M88.7,56.8c0,5.5-4.5,10.1-10.1,10.1c-5.6,0-10.1-4.6-10.1-10.1c0-5.6,4.5-10.1,10.1-10.1C84.2,46.7,88.7,51.3,88.7,56.8z};
  }
}

\newcommand\orcidicon[1]{\href{https://orcid.org/#1}{\mbox{\scalerel*{
\begin{tikzpicture}[yscale=-1,transform shape]
\pic{orcidlogo};
\end{tikzpicture}
}{|}}}}

\usepackage{hyperref} 

\begin{document}
\title{Defending Democracy: Using Deep Learning to Identify and Prevent Misinformation
}
\titlerunning{Defending Democracy}        
\author{Anusua Trivedi \orcidicon{0000-0002-1315-2113} \and
        Alyssa Suhm\and
        Prathamesh Mahankal\and
        Subhiksha Mukuntharaj\and
        Meghana D. Parab\and
        Malvika Mohan\and
        Meredith Berger\and
        Arathi Sethumadhavan\and
        Ashish Jaiman\and
        Rahul Dodhia
}
\authorrunning{Trivedi et. al.} 
\institute{Anusua Trivedi \at Microsoft, \email{antriv@microsoft.com},
           \and
           Alyssa Suhm \at Microsoft, \email{v-alsuhm@microsoft.com}
           \and
           Prathamesh Mahankal \at University of Washington, \email{psm1695@uw.edu}
           \and
           Subhiksha Mukuntharaj \at University of Washington, \email{subhi@uw.edu}
           \and
           Meghana D. Parab \at University of Washington, \email{megs1695@uw.edu}
           \and
           Malvika Mohan \at University of Washington, \email{pmmohan05@uw.edu}
           \and
           Meredith Berger \at Microsoft, \email{Meredith.Berger@microsoft.com}
           \and
           Arathi Sethumadhavan \at Microsoft, \email{Arathi.Sethumadhavan@microsoft.com}
           \and
           Ashish Jaiman \at Microsoft, \email{ashish.jaiman@microsoft.com}
           \and
           Rahul Dodhia \at Microsoft, \email{radodhia@microsoft.com}
}
\maketitle
\begin{abstract}
The rise in online misinformation in recent years threatens democracies by distorting authentic public discourse and causing confusion, fear, and even, in extreme cases, violence. There is a need to understand the spread of false content through online networks for developing interventions that disrupt misinformation before it achieves virality. Using a Deep Bidirectional Transformer for Language Understanding (BERT) and propagation graphs, this study classifies and visualizes the spread of misinformation on a social media network using publicly available Twitter data. The results confirm prior research around user clusters and the virality of false content while improving the precision of deep learning models for misinformation detection. The study further demonstrates the suitability of BERT for providing a scalable model for false information detection, which can contribute to the development of more timely and accurate interventions to slow the spread of misinformation in online environments.
\keywords{democracy \and misinformation \and social media \and virality \and deep learning \and transformers}
\end{abstract}

\section{Introduction}
A collective recognition of authoritative sources of credible information is essential to the functioning of a democracy. People have access to more information through the Internet than ever before, and nefarious actors use this hyper-connected digital world to spread disinformation at unprecedented speed and scale. From the individual to the societal level, misinformation has real-world impacts, including increasing polarization, reinforcing societal divisions, and even inciting violence. Misinformation distorts public discourse and disrupts the authentic exchange of ideas and opinions. 

Types of problematic information exist on a spectrum based on falseness and harm~\cite{RefE}. Misinformation, while false, is not intended to deceive or cause harm. In contrast, disinformation is false and created or distributed with the intent to deceive or cause harm. The general motive to spread disinformation is to damage an entity's reputation, mislead the readers, or gain from sensationalism. It is seen as one of the greatest threats to democracy, open debate, and free and modern society~\cite{RefL}.

Despite these distinctions in taxonomy, some authors use “misinformation” as an umbrella term to refer to all types of false information. Brennen et al.~\cite{RefF} states that they use “misinformation” to refer broadly to any false information because it is hard to assess if false content was created to deceive or by mistake. They focus on false information, including both misinformation and disinformation. This paper follows this precedent in using “misinformation," as the datasets used to train the model focus only on whether the content is false, regardless of intent.

Prior research confirms that Americans across the political spectrum are concerned about false information~\cite{RefA}. According to Pew Research Center, 63\% of Americans surveyed report that they do not trust the news they encounter on social media, despite 67\% of respondents also acknowledging that they use social media to get news on a regular basis~\cite{RefB}. Trust in information is shaped by social norms and individual belief systems, and once false information gains acceptance, it can be highly resistant to correction~\cite{RefB1}. Additionally, studies show that false information travels significantly farther, faster, deeper, and more broadly than credible information on social media platforms~\cite{RefC}. Previous authors have hypothesized that this may be due to the novelty of false information and the emotional response it often evokes, such as surprise, fear, or disgust~\cite{RefC}. Current responses to misinformation on social media platforms are often reactive, taking place after the content has been distributed widely and after the damage has already largely been done. Unfortunately, even when widely discredited, exposure to misinformation can create lingering effects on peoples' attitudes and perceptions~\cite{RefD}.

Misinformation is increasingly being shared via social media platforms like Twitter, YouTube, and Facebook~\cite{RefK}. These platforms offer the general population a forum to share their opinions and views in a raw and unedited fashion. Some news articles hosted or shared on social media platforms have more views than direct views from the media outlets’ platforms. Researchers studied the velocity of misinformation and concluded that tweets containing false information reach people on Twitter six times faster than truthful tweets~\cite{RefL}. 

There is a need better to understand the spread of false content through online networks to develop interventions that disrupt misinformation before it achieves virality. According to Granik et al.~\cite{RefG}, artificial intelligence performs better to detect misinformation. In this paper, we utilize a deep learning model to predict if the content is false. We also use visualizations of how that content spreads through social media over time as an approach to detecting the virality of false content. 

\section{Prior Work}
Detecting misinformation is one of the most challenging tasks for a human being. Researchers have explored supervised machine learning models to detect misinformation. Abdullah-All-Tanvir et al.~\cite{RefH} used SVM and Naïve Bayes classifiers for detecting false content. Rahman et al. ~\cite{RefI} found that the Logistic regression is a better classifier model. According to Ahmed et al. ~\cite{RefJ}, false information has a significant impact on the political situation of society. They use a Naïve Bayes classifier to detect false information from Facebook posts. Most of this work is based on different handheld features, which can capture linguistic and psychological causes. However, these features failed to classify text well, which limits the generalized performance of the models. Thus researchers started exploring neural network-based models that learn document-level representation to discover deception text. Neural network models have been shown to learn semantic representations for NLP tasks successfully.  Bambu et al. ~\cite{RefM} have recently been able to show the use of graph-based methods and deep learning approaches like CNN in combating false information on social media. In this paper, we focus on the application of Deep Bidirectional Transformers for Language Understanding (BERT) ~\cite{RefN} to the problem of misinformation detection.
\begin{figure*}
\centering
\captionsetup{justification=centering}
\fbox{\includegraphics[width=0.75\textwidth]{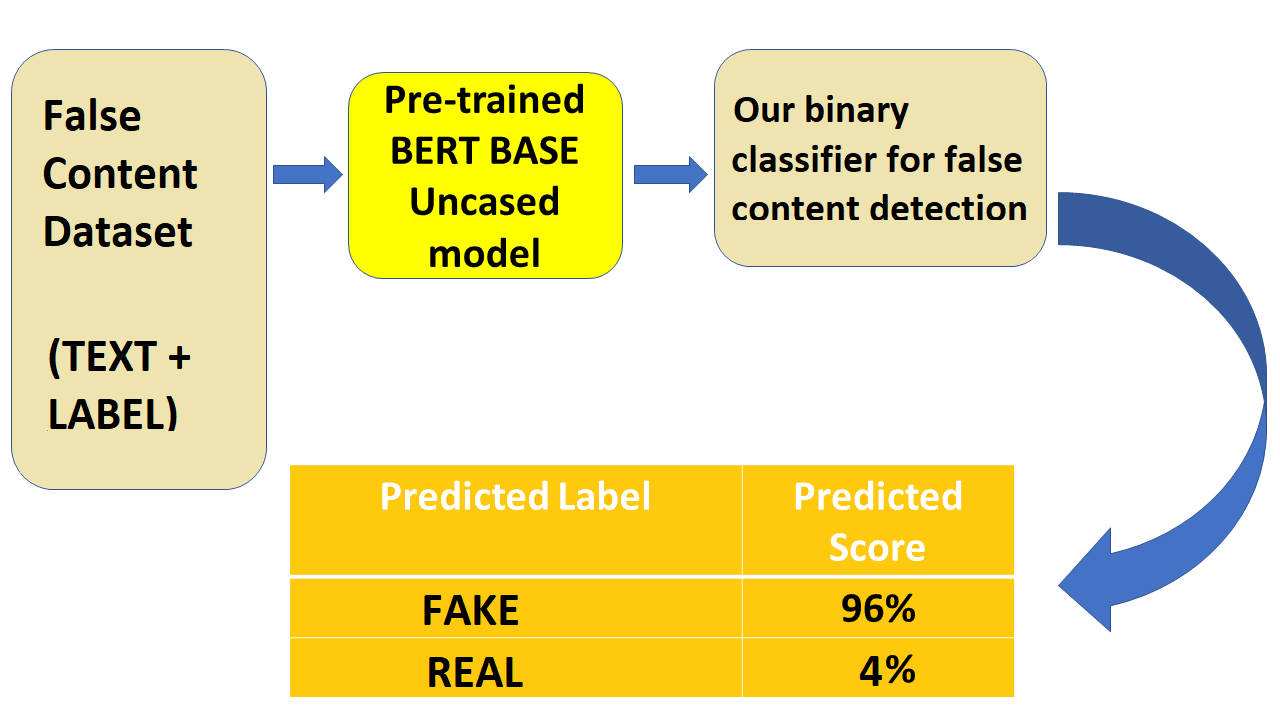}}
\caption{BERT binary classification model for Misinformation Detection}
\label{fig:1} 
\end{figure*}

\begin{figure*}
\centering
\captionsetup{justification=centering}
\fbox{\includegraphics[width=0.75\textwidth]{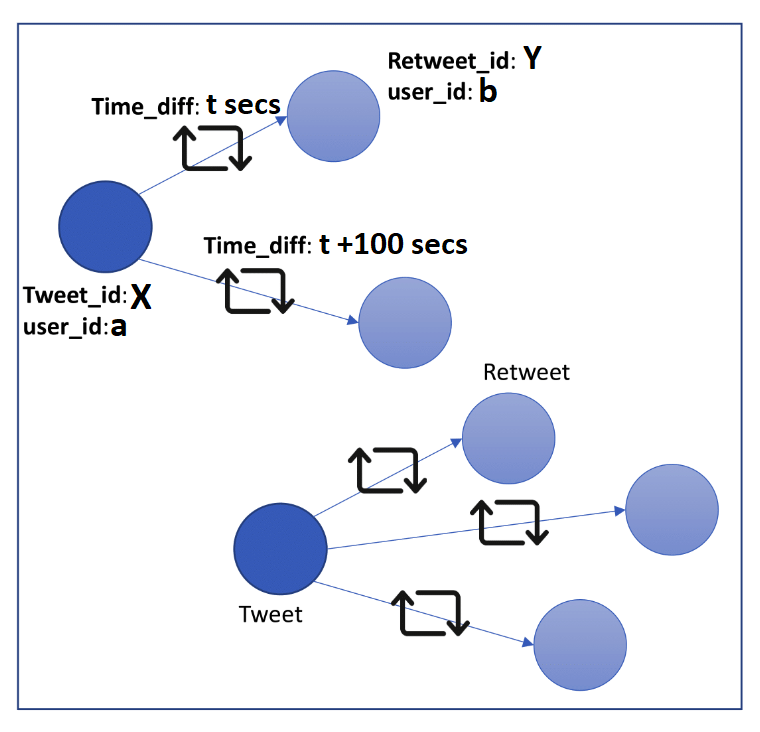}}
\caption{Example of a propagation graph as described in~\cite{RefV1}}
\label{fig:2} 
\end{figure*}

\begin{figure*}
\centering
\captionsetup{justification=centering}
\fbox{\includegraphics[width=0.75\textwidth]{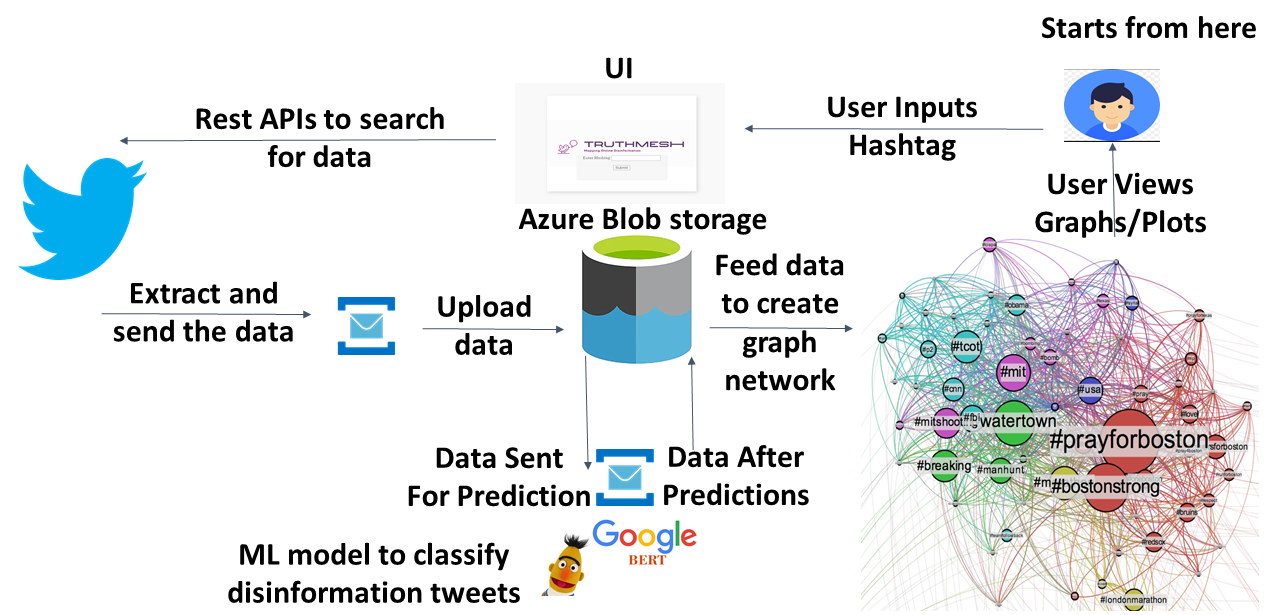}}
\caption{Using binary misinformation classification result to detect virality trend of hashtags}
\label{fig:3} 
\end{figure*}

\begin{figure*}
\centering
\captionsetup{justification=centering}
\fbox{\includegraphics[width=0.75\textwidth]{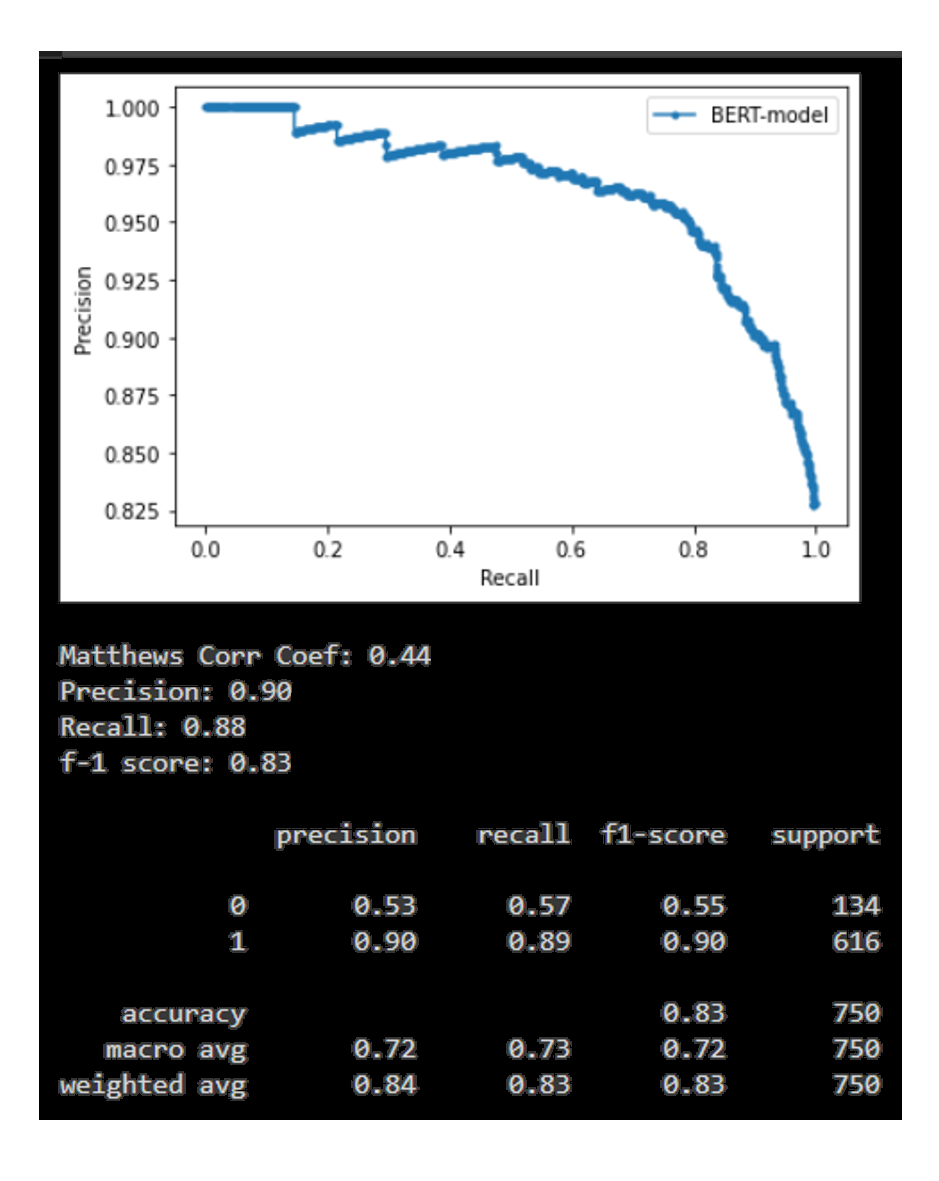}}
\caption{Precision-Recall curve and F1-Score of our model on validation dataset}
\label{fig:4} 
\end{figure*}

\section{Data}
As an initial stage towards building an automated solution for identifying misinformation, we curated datasets from different open source labeled databases curated by experts. The dataset used to train the machine learning model was created by merging text content (independent variable) and fake/real labels (dependent variable) from different public data sources. Below is the list of the data sources used:
\begin{itemize}
    \item{\textbf{LIAR}}: LIAR~\cite{RefN1} is a publicly available dataset for fake news detection. The LIAR dataset consists of 12.8K human-labeled short statements from Politifact.com’s API, and a Politifact.com editor for its truthfulness evaluates each statement. This dataset can be used for fact-checking as well. Notably, this new dataset is an order of magnitude larger than the previously largest public fake news datasets of similar type. 
    \item{\textbf{ISOT Fake News}}: The ISOT Fake News dataset~\cite{RefN2} is a compilation of several thousand fake news and truthful articles obtained from different legitimate news sites and sites flagged as unreliable by Politifact.com.
    \item{\textbf{Horne 2017 Fake News Data}}: Introduced by Horne et al. ~\cite{RefN3}, this dataset consists of two independent news datasets:
        \subitem{\textbf{Buzzfeed Political News Data}}: This news dataset is analyzed by Craig Silverman et al. ~\cite{RefN4}, where they use the keyword search on the content analysis tool “BuzzSumo” to find news stories and post the analysis. The authors collect the body text and body title of all articles and use the ground truth set by Buzzfeed as the actual ground truth. This dataset has fewer clear restrictions on the ground truth, including opinion-based real stories and satire-based fake stories. We manually filter this data to contain only “hard” news stories and “malicious” fake news stories for our paper.
        \subitem{\textbf{Random Political News Data}}: This consists of randomly collected news from Business Insider’s “Most Trusted” list and Zimdars 2016 Fake news list. The “Real” news came from Wall Street Journal, the Economist, BBC, NPR, ABC, CBS, USA Today, the Guardian, NBC, and the Washington Post. The “Satire” news came from The Onion, Huffington Post Satire, Borowitz Report, the Beaverton, Satire Wire, and Faking News. The “Fake” news came from Ending The Fed, True Pundit, DC Gazette, Liberty Writers News, Before its News, InfoWars, Real News Right Now. Our paper uses only the “Real” and “Fake” labels and text from this dataset.
    \item{\textbf{Russian Troll tweets}}: This dataset is compiled and published by Linvill et al. ~\cite{RefN5}. As part of the House Intelligence Committee investigation into how Russia may have influenced the 2016 US Election, authors released almost 3000 Twitter accounts believed to be connected to Russia’s Internet Research Agency, a company known for operating social media troll accounts.
    \item{\textbf{FakeNewsNet}}: Shu et al. ~\cite{RefN6} constructed and publicized a multi-dimensional data repository, which contains two datasets with news content, social media context, and spatiotemporal information. The dataset is constructed using an end-to-end FakeNewsTracker~\cite{RefN7} system.
    \item{\textbf{UTK Machine Learning Club}}: UTK Machine Learning Club curated a dataset to identify if an article might be fake news ~\cite{RefN8}. 
    \item{\textbf{Kaggle Fake News Detection}}: Kaggle user Jruvika curated a dataset of fake news and open-sourced it for public use ~\cite{RefN9}. 
    \item{\textbf{NBC Election Troll Tweets 2016}}: This data set consists of 11.5k expanded news links in set of 36.5k 2016 election troll tweets~\cite{RefN10}.
    \item{\textbf{Fake News on Twitter 2016 viral tweets}}: Longa et al. ~\cite{RefN11} curated a collection of tweets that went viral on 2016 US election day (up until March 2017). Viral tweets are those that achieved the 1000-retweet threshold during the collection period. We queried Twitter’s streaming API using the hashtags \#MyVote 2016, \#ElectionDay, \#electionnight, and the user handles @realDonaldTrump and @HillaryClinton.
\end{itemize}
The final combined dataset has 98532 text content and binary ‘Fake/Real’ (1=fake, 0=real) label pairs. All the above datasets were transformed to make labels binary wherever needed. For example, we labeled the rows as ‘Real’ for the LIAR dataset if the content was labeled ‘true’ or ‘mostly-true’ and ‘Fake’ if the content was labeled ‘pants-on-fire’ or ‘false’. Content labeled ‘half-true’ or ‘barely-true’ was not included.
\section{Method}
Text Classification is a long-standing problem in NLP, and the community has introduced several paradigms that differ from each other in type and approach. In this paper, we are particularly interested in the context-aware misinformation classification paradigm. The label for each text snippet can be obtained by referring to its accompanying context (paragraph or a list of sentences). In this paper, we approach the problem in two steps:
\begin{itemize}
\item Build a transformer-based classifier for real and false information prediction.
\item Use this prediction model to extract features from the propagation graphs to further analyze how misinformation propagates on social media (e.g. Twitter).
\end{itemize}

\subsection{\textbf{Fine-tuning BERT for binary misinformation classification}}
Research in the field of using pre-trained deep learning models has resulted in a massive leap in state-of-the-art results for text classification. ELMo~\cite{RefO}, ULMFiT~\cite{RefP} and OpenAI Transformer~\cite{RefQ} are some of the popular models which have shown immense success in the text classification task. All these models allow us to pre-train an unsupervised language model on a large corpus of data such as all Wikipedia articles and then fine-tune these pre-trained models on downstream tasks. More recent methods mostly employ the Transformer architecture~\cite{RefQ1} as their base model design, which does not process sequentially like RNNs and instead relies entirely on the attention mechanism, e.g., BERT~\cite{RefN}, GPT-2~\cite{RefQ2}, XLNet~\cite{RefQ3}, MegatronLM~\cite{RefQ4}, and Turing-NLG~\cite{RefQ5}. Transformer models have complex architecture consisting of multiple parts such as embedding layers, self-attention layers, and feed-forward layers. These are powerful models with an improved state of the art for different NLP tasks by significant margins.

Modern deep neural networks often significantly benefit from transfer learning. Deep convolutional neural network (CNN)~\cite{RefR} trained on a large image classification dataset such as ImageNet~\cite{RefS} have proved to help initialize models on other computer vision tasks~\cite{RefT}. Transfer learning for a new purpose-specific model, using the pre-trained neural network as the basis, has shown immense success. 
In recent years, similar transfer learning techniques have been useful for many natural language tasks. We use the smaller BERT-Base-uncased model as the base model for this paper. The BERT-Base-uncased model has 12 attention layers and uses the word-piece-tokenizer~\cite{RefU}, which converts all text to lowercase. We modify the Bert-For-Sequence-Classification class in BERT GitHub~\cite{RefV} for multi-label classification. We use binary-cross-entropy-with-logits~\cite{RefW} as loss function for binary classification task instead of the standard cross-entropy loss used by the BERT model. Binary-cross-entropy loss allows our model to assign independent probabilities to the labels. Figure~\ref{fig:1} explains our multilabel classification pipeline. The training logic is identical to the one provided in classifier.py in~\cite{RefV}. We train the model for ten epochs with a batch size of 32 and sequence length as 256. The learning rate is kept to 3e-5, as recommended for fine-tuning in the original BERT paper. We do not use the precision FP16 technique as the binary-cross-entropy-with-logits loss function does not support FP16 processing.

\subsection{\textbf{Using binary misinformation classification model to create Propagation Graphs in Twitter}}
As described by Marion et al.~\cite{RefV1}, propagation graphs are derived from the set of tweets and re-tweets corresponding to a piece of information (e.g. hashtag). 
Let's define that set as T. A node can be of two types -
\begin{itemize}
    \item{\textbf{A tweet node}}: The node stores the tweet and its associated user. A tweet belongs to a news graph if it contains the keywords extracted from the news article's headline.
    \item{\textbf{A re-tweet node}}: The node stores the re-tweet and its associated user. All re-tweets of a tweet node are present in the graph. 
\end{itemize}

Edges are drawn between a tweet and its re-tweets. Edges contain a time weight that corresponds to the time difference between the tweet and re-tweet publish times. Let the set of edges of the graph be E. 

A news graph comprises non-connected sub-graphs where each sub-graph comprises a tweet and its associated re-tweets. Then we can define a news graph as N = (T,E).

It is important to note that Twitter is designed in such a way that a re-tweet of a re-tweet will point back to the original tweet. Hence, the depth of the graph is never more than 1. Figure~\ref{fig:2} shows an example of a propagation graph.

We create a Twitter Propagation graph to enhance the visibility and symbolic power of social media information. Here we try to examine how characteristics of hashtags drove the misinformation virality during a social networking event. We extract data pertinent to a hashtag, send it to our machine learning model for misinformation detection, and then users are clustered together with the help of community detection algorithms from Louvain~\cite{Refa} and InfoMap~\cite{Refb}. The whole architecture is defined in Figure ~\ref{fig:3}. Through propagation graphs, we can identify popular hashtag types and examined hashtag co-occurrence patterns. The graph visualization of this clustered output helps us detect re-tweets and track hashtag virality of misinformation.

\section{Evaluation of our model}
\subsection{\textbf{Classifier Outcome}}
Out of the 98,532 text content, we used 80\% for training and 20\% for evaluating the model. The validation dataset consists of 19,706 records and the evaluation of our model on that held-out dataset had an F1 score of 84.43\% (Precision = 0.90, Recall = 0.89, Mathews Corr Coef = 0.44). Figure~\ref{fig:4} shows the precision-recall curve and confusion matrix on our validation dataset.

\begin{figure*}
\centering
\captionsetup{justification=centering}
\fbox{\includegraphics[width=\textwidth]{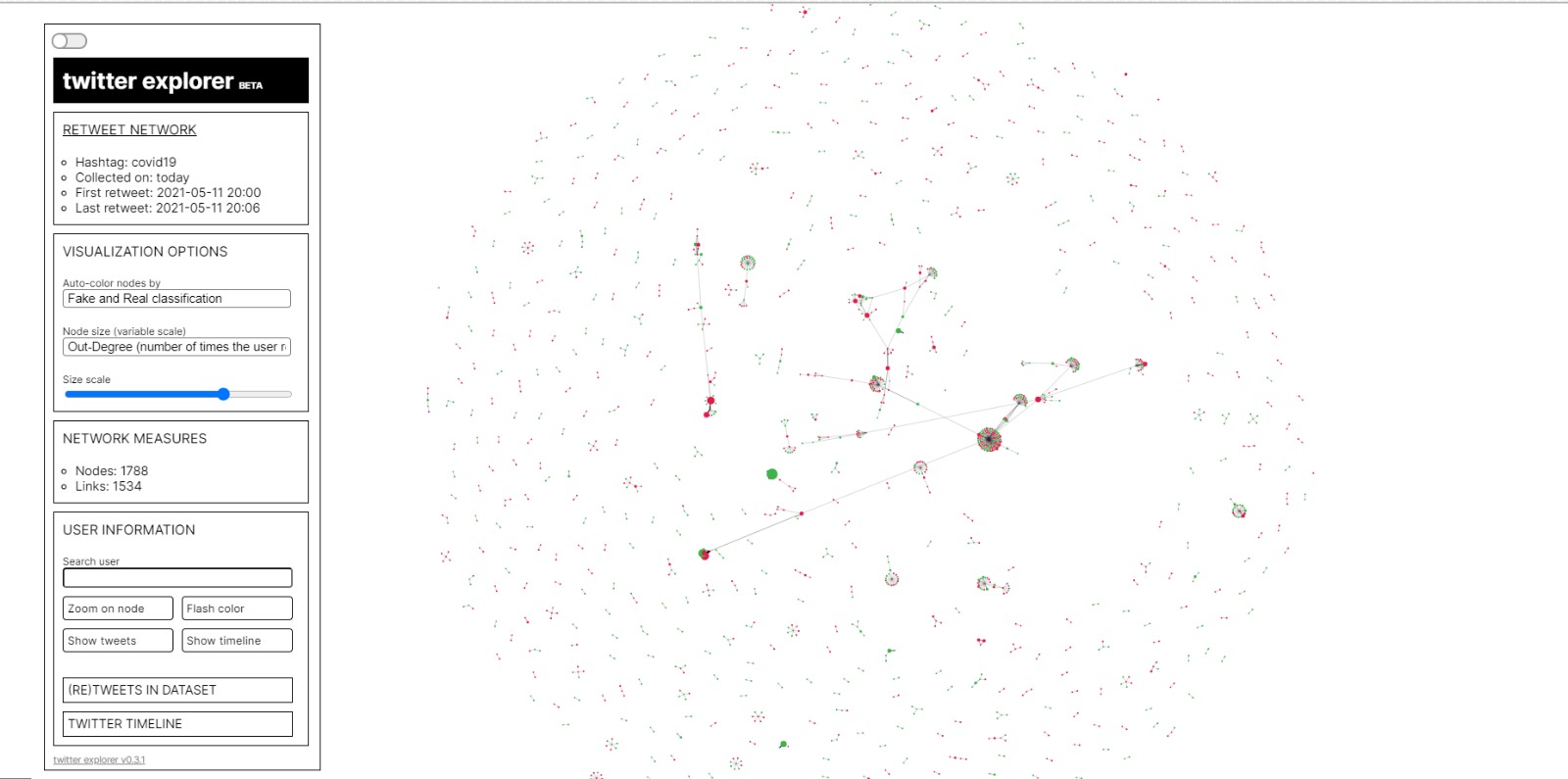}}
\caption{Tweet retweet interaction network graph. Here the node size is scaled according to out degree score. The nodes have been recolored to reflect fake and real tweets.}
\label{fig:5} 
\end{figure*}

\begin{figure*}
\centering
\captionsetup{justification=centering}
\fbox{\includegraphics[width=\textwidth]{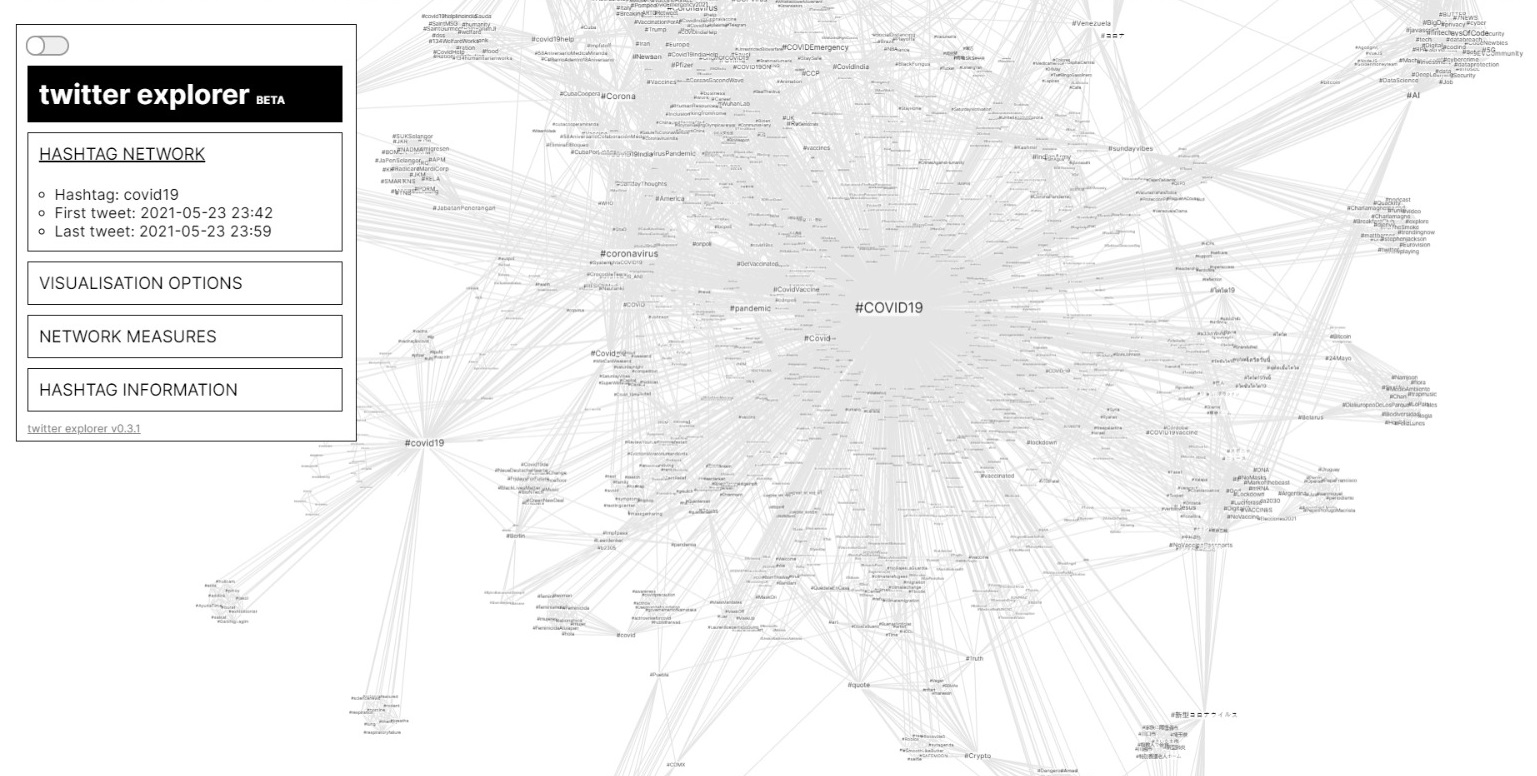}}
\caption{Hashtag network graph for tweets collected with hashtag \#covid19. The graph shows the frequency of occurrence of two hashtags in the same tweet. The size of the hashtag shows the number of times it has been used.}
\label{fig:6} 
\end{figure*}

\subsection{\textbf{Propagation Graph Analysis}}
The Propagation Graph analysis for a particular hashtag is shown in Figure~\ref{fig:5} and Figure~\ref{fig:6}. Users have been clustered together with the help of community detection algorithms from Louvain~\cite{Refa} and InfoMap~\cite{Refb}. The tweet-retweet interaction graph in Figure~\ref{fig:5} shows nodes as users and links between nodes if a user has re-tweeted. Users who have re-tweeted are placed closer to each other due to the force-directed layout. In Figure~\ref{fig:6}, each node of the hashtag network is a hashtag used by a user. A link is present between hashtags if they appear in the same tweet. Hashtag text size depicts how often it is used. These graph visualizations give more context about what the users are interacting with. These are also spread out in a force-directed layout. 

For both the graphs, we can obtain information about the number of nodes, links, user information, first and last date of the tweet. Based on the number of followers, the node size can be changed. This helps to show how influential the user sharing real or false information could be. The code for both these visualizations is written partly by the visualizer in Twitter explorer~\cite{Refc}. Python implementation of the igraph library~\cite{Refd} is used for making the graph networks. D3.js~\cite{Refe} is also used for creating the interactive networks and force-graph.

\section{Discussion}
BERT misinformation classification provides a significant step forward in applying deep learning to detecting and slowing the spread of false information online. The results confirm prior work indicating that misinformation content on a particular topic typically originates from a small number of core users. Further, temporal information on trending topics indicates that virality is usually achieved within several hours to days. These initial findings suggest that BERT may perform with a higher level of precision than prior methods, contributing to the continued refinement of AI-enabled misinformation detection. However, subsequent testing is needed to evaluate the model's performance on unfamiliar datasets. 

This research contributes to a better academic understanding of the spread of information online and the development of better technological interventions to counter misinformation. Interventions such as outright content removal or user account takedown are controversial and challenging to enforce equitably, especially given that the window of time to respond to a viral misinformation campaign is narrow. Therefore, these results and future research could help better inform other response options, including algorithmic interventions or labeling and contextualization.  

Future research should seek to apply this approach to data from other social media platforms and misinformation content in non-English languages. An additional challenge requiring further study is classifying information that is partially true but mischaracterized or mis-contextualized. Finally, as those trying to detect misinformation will always face an asymmetric disadvantage to those trying to spread it, future technology solutions will likely shift towards media authentication and provenance tools that enable verification at the point of origin and throughout the lifecycle of a piece of content.

Misinformation threatens authentic democratic discourse and the shared conceptualization of reality that underpins healthy societies. While upholding freedom of speech and expression, it is also important to maintain an individual's right to access accurate, truthful information. Identifying and mapping the spread of misinformation through online networks will help those seeking to curb its impact develop more accurate, equitable, and transparent interventions.

\section*{Declarations}
\subsection*{Funding: Not applicable}
\subsection*{Conflict of interest: The authors declare that they have no conflict of interest.}
\subsection*{Availability of data and material: All data is open source.}
\subsection*{Code availability: \url{https://github.com/prathameshmahankal/Fake-News-Detection-Using-BERT}}
\subsection*{Ethics approval: Not applicable}
\subsection*{Consent to participate: The authors declare that they consent to participate.}
\subsection*{Consent for publication: The authors declare that they consent to publication.}

\end{document}